\documentclass[english,10pt,aps,pra,twocolumn,groupedaddress,floatfix]{revtex4-1}

\usepackage{amsmath}
\usepackage{amssymb}
\usepackage{amsfonts}
\usepackage{bm}
\usepackage{bbm}
\usepackage{epsfig}
\usepackage{grffile}
\usepackage{graphics}

\usepackage[usenames,dvipsnames]{color}
\definecolor{dblue}{rgb}{0,0.1,.6}

\usepackage[colorlinks=true,citecolor=dblue,linkcolor=dblue,urlcolor=dblue]{hyperref}
\usepackage[all]{hypcap}

\newcommand{\bra}{\langle}
\newcommand{\ket}{\rangle}
\newcommand{\bs}{\boldsymbol}
\newcommand{\Tr}{\operatorname{Tr}}

\newcommand{\mr}[1]{\mathrm{#1}}

\newcommand{\hH}{\hat{H}}
\newcommand{\hh}{\hat{h}}
\newcommand{\hS}{\hat{S}}
\newcommand{\hA}{\hat{A}}
\newcommand{\hB}{\hat{B}}
\newcommand{\hX}{\hat{X}}
\newcommand{\hP}{\hat{P}}
\newcommand{\hQ}{\hat{Q}}
\newcommand{\hsigma}{\hat{\sigma}}
\newcommand{\hrho}{\hat{\rho}}
\newcommand{\diag}{\mathrm{diag}}

\newcommand{\hSvec}{\hat{\bs{S}}}

\newcommand{\vn}{{\bs{n}}}
\newcommand{\vez}{{\bs{e}_z}}
\newcommand{\vl}{{\bs{l}}}
\newcommand{\vm}{{\bs{m}}}

\newcommand{\mc}[1]{\mathcal{#1}}

\newcommand{\trunc}{\text{trunc}}

\usepackage{xr}

\newcommand{\duke} {Department of Physics, Duke University, Durham, North Carolina 27708, USA}

\newcommand{\Title} {Low-energy physics of isotropic spin-1 chains in the critical and Haldane phases}
\newcommand{\Authors}
{
\author{Moritz Binder}
\affiliation{\duke}
\author{Thomas Barthel}
\affiliation{\duke}
}
\newcommand{\Date} {July 9, 2020}

\begin{document}

\title{\Title}
\Authors

\begin{abstract}
Using a matrix product state algorithm with infinite boundary conditions, we compute high-resolution dynamic spin and quadrupolar structure factors in the thermodynamic limit to explore the low-energy excitations of isotropic bilinear-biquadratic spin-1 chains. Haldane mapped the spin-1 Heisenberg antiferromagnet to a continuum field theory, the non-linear sigma model (NL$\sigma$M). We find that the NL$\sigma$M fails to capture the influence of the biquadratic term and provides only an unsatisfactory description of the Haldane phase physics. But several features in the Haldane phase can be explained by non-interacting multi-magnon states. The physics at the Uimin-Lai-Sutherland point is characterized by multi-soliton continua. Moving into the extended critical phase, we find that these excitation continua contract, which we explain using a field-theoretic description. New excitations emerge at higher energies and, in the vicinity of the purely biquadratic point, they show simple cosine dispersions. Using block fidelities, we identify them as elementary one-particle excitations and relate them to the integrable Temperley-Lieb chain.
\end{abstract}

\date{\Date}
\maketitle

\section{Introduction}
The most general model for spin-1 chains with isotropic nearest-neighbor interactions is the bilinear-biquadratic Hamiltonian
\begin{equation} \label{eq:H_blbq}
	\hH_\theta = \sum_i\left[\cos\theta (\hSvec_i \cdot \hSvec_{i+1}) + \sin\theta (\hSvec_i \cdot \hSvec_{i+1})^2\right],
\end{equation}
where the angle $\theta\in[-3\pi/4,5\pi/4)$ parametrizes the ratio of the two couplings. It describes quasi-one-dimensional quantum magnets like CsNiCl$_3$ \cite{buyers_1986,tun_1990,zaliznyak_2001}, Ni(C$_2$H$_8$N$_2$)$_2$NO$_2$ClO$_4$ (NENP) \cite{ma_1992,regnault_1994},
or LiVGe$_2$O$_6$ \cite{millet_1999,lou_2000}, and can be realized with cold atoms in optical lattices \cite{yip_2003,imambekov_2003,garcia_ripoll_2004}. Depending on $\theta$, the ground state can be in one of several interesting quantum phases. In addition to a ferromagnetic ($\pi/2 < \theta$) and a gapped dimerized phase ($-3\pi/4 < \theta < -\pi/4$) \cite{barber_1989,kluemper_1989,xian_1993,laeuchli_2006}, the model features the gapped Haldane phase ($-\pi/4 < \theta <\pi/4$) \cite{haldane_1983_pla,haldane_1983_prl,affleck_1989,schmitt_1998} characterized by symmetry-protected topological order \cite{gu_2009,pollmann_2012}, and an extended critical phase ($\pi/4 \leq \theta < \pi/2$) \cite{uimin_1970,lai_1974,sutherland_1975,laeuchli_2006,fath_1991,fath_1993,itoi_1997}.

While the groundstate phase diagram has been studied extensively, much less is known about the low-energy dynamics that we address in this paper. We use a recently introduced algorithm \cite{binder_2018} based on the density matrix renormalization group (DMRG) \cite{white_1992,white_1993,schollwoeck_2005} and the time evolution of matrix product states (MPS) \cite{vidal_2004,white_2004,daley_2004} with infinite boundary conditions \cite{mcculloch_2008,phien_2012} to compute high-resolution dynamic structure factors
\begin{equation} \label{eq:dsf_def}
	S(k, \omega) = \sum_x e^{-ikx} \int\!\mr{d}t\,e^{i\omega t} \bra\psi|\hA_x(t) \hB_0(0)|\psi\ket
\end{equation}
in the thermodynamic limit. Here, $\hX(t) := e^{i\hH t}\hX\, e^{-i\hH t}$ and $|\psi\ket$ is the ground state. As the model is SU(2)-symmetric and the ground states are singlets, it follows from the Wigner-Eckart theorem that there are only two independent structure factors for one-site operators -- the \emph{spin} structure factor $S^{zz}(k, \omega)$ where $\hA=\hB=\hS^z$, and the \emph{quadrupolar} structure factor $S^{QQ}(k, \omega)$ where $\hA=\hB=\hQ=\diag(1/3, -2/3, 1/3)$. As the ground states of interest are singlets ($S_{\mr{tot}}=0$), selection rules imply that these structure factors probe excitations with total spin quantum numbers $S_{\mr{tot}}=1$ and $2$, respectively.
They allow us to elucidate the low-energy dynamics of the model. To this end, we compare the numerical results to Bethe ansatz and field-theoretical treatments. In this work, we focus on the Haldane phase and the extended critical phase, which have the most interesting physics.

The structure factors provide a direct link between theory and experiment, as they can be measured by inelastic neutron scattering  experiments \cite{nagler_2020}. An important example is their precise measurement for the spin-1 antiferromagnet CsNiCl\textsubscript{3} \cite{zaliznyak_2001}. Measurements of the magnetic susceptibility of the vanadium oxide LiVGe\textsubscript{2}O\textsubscript{6} substantiated the presence of biquadratic interactions \cite{millet_1999,lou_2000}, and our results provide a prediction for future measurements of the full dynamic structure factor for this material. In addition, experimental setups with cold atoms in optical lattices have been proposed which allow for the realization of the model \eqref{eq:H_blbq} with control on the parameter $\theta$ \cite{yip_2003,imambekov_2003,garcia_ripoll_2004}. For these systems, dynamic structure factors can be measured using inelastic scattering of photons \cite{stenger_1999,stewart_2008,dao_2007,rey_2005,clement_2010,ernst_2010,landig_2015}.

\section{Haldane phase}
For the Heisenberg antiferromagnet ($\theta=0$) with vanishing biquadratic term, Haldane mapped the model to a continuum field theory, the $O(3)$ non-linear sigma model (NL$\sigma$M) \cite{haldane_1983_pla,haldane_1983_prl}, by restricting to the most relevant low-energy modes at momenta $k=0$ and $\pi$. The mapping becomes exact in the limit of large spin $S\to\infty$. The NL$\sigma$M is integrable and predicts an energy gap to the lowest excited states, which is known as the Haldane gap. This is at the heart of the famous Haldane conjecture, according to which the physics of integer and half-integer antiferromagnetic spin chains is fundamentally different.

Based on the NL$\sigma$M description, one expects the lowest excited states to be a triplet of single-magnon states at momentum $k=\pi$. The single-magnon dispersion near $k=\pi$ is predicted to be of the form 
\begin{equation} \label{eq:nlsm_dispersion}
	\varepsilon_\text{NL$\sigma$M}(k)=\sqrt{\Delta^2 + v^2 (k-\pi)^2}
\end{equation}
with the energy gap $\Delta$, and the spin-wave velocity $v$. Correspondingly, the onset of a two-magnon continuum at $(k,\omega)=(0,2\Delta)$ and of a three-magnon continuum at $(k,\omega)=(\pi,3\Delta)$ are predicted, and the contributions of these continua to the dynamic structure factors have been computed for the NL$\sigma$M \cite{affleck_1992,horton_1999,essler_2000,essler_2005}.
\begin{figure}[t]
 \begin{center}
  \includegraphics[width=\columnwidth]{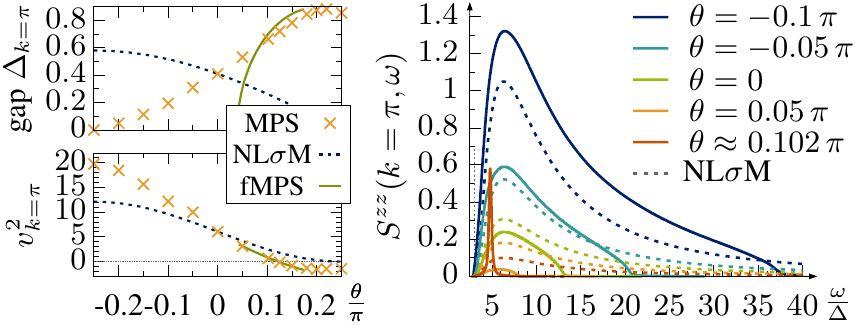}
 \end{center}
 \caption{Comparison of the NL$\sigma$M and fMPS predictions with numerical results. Left: $\theta$-dependence of the excitation gap $\Delta$ and the squared spin-wave velocity $v^2$ at $k=\pi$. The NL$\sigma$M values have been scaled to match numerics at $\theta=0$. Right: Dynamic structure factor $S^{zz}(k=\pi, \omega)$ versus the NL$\sigma$M three-magnon continuum. For the comparison, the latter has been scaled by matching the single-magnon weights and multiplying a factor of 4.}
 \label{fig:nlsm}
\end{figure}

To study the applicability of the NL$\sigma$M for the physics in the Haldane phase, we include the biquadratic term from the Hamiltonian \eqref{eq:H_blbq} in the mapping to the field theory. Details are provided in Appendix~\ref{sec:mapping_to_NLSM}. In the end, this boils down to evaluating the matrix element of the biquadratic interaction with respect to spin-coherent states. Using the fact that higher-order terms vanish in the continuum limit, we find that the biquadratic term does not change the form of the resulting action. Its effect is a renormalization of the coupling constant $J$ such that $J(\theta) = J(0)(\cos\theta - \sin\theta)$. As long as the biquadratic term is sufficiently small, the identification of the relevant degrees of freedom and the further derivations remain valid. Thus, one would expect the physics to be unchanged for a region around $\theta \approx 0$ with the renormalization leading to a $\theta$ dependence of the gap and the spin-wave velocity with $\Delta,v\propto\cos\theta - \sin\theta$.

Surprisingly, these predictions strongly disagree with our numerical data as shown in Fig.~\ref{fig:nlsm}. For the simulations, we employ a fourth-order Lie-Trotter-Suzuki decomposition \cite{trotter_1959,suzuki_1976,barthel_2019} with time step $\tau=0.1$ and truncate components with Schmidt coefficients $\lambda_k<\lambda_\trunc$ and truncation thresholds in the range $\lambda^2_\trunc \sim 10^{-10} - 10^{-8}$, depending on $\theta$.

\begin{figure}[t]
 \begin{center}  
  \includegraphics[width=\columnwidth]{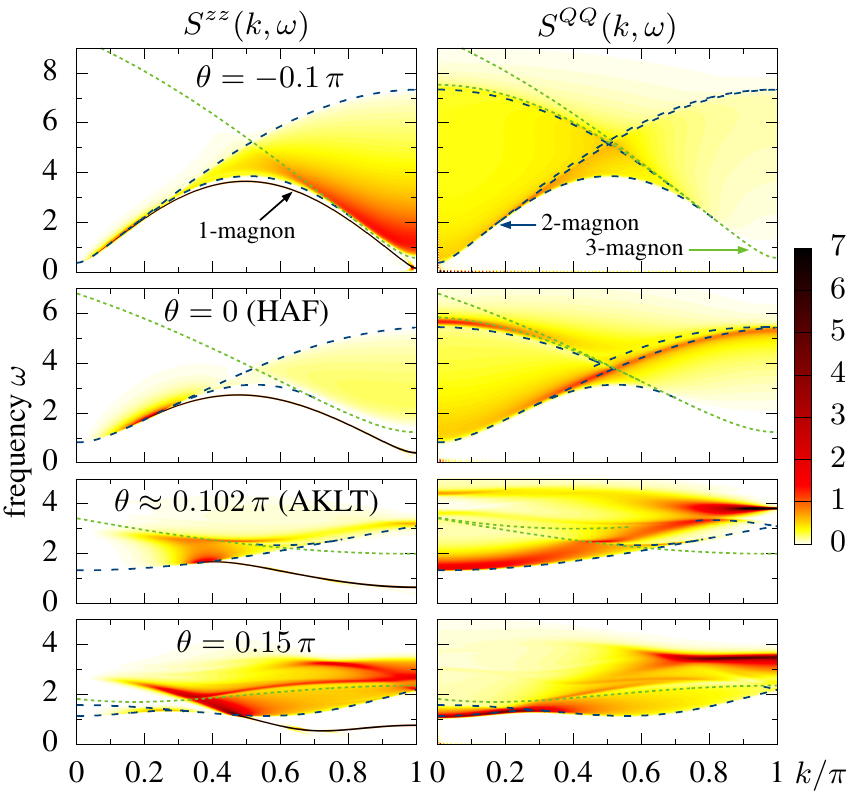}
 \end{center}
 \caption{The dynamic spin and quadrupolar structure factors in the Haldane phase. Dashed lines indicate thresholds for two- and three-magnon continua in the non-interacting approximation.}
 \label{fig:dsf_Haldane}
\end{figure}

While the NL$\sigma$M predicts a \emph{decreasing} gap when we increase $\theta$, the actual gap \emph{increases}. For the spin-wave velocity, the trend suggested by the NL$\sigma$M seems correct at first sight. However, after crossing the Affleck-Kennedy-Lieb-Tasaki (AKLT) point $\theta=\arctan(1/3)\approx 0.1024\pi$ \cite{aklt_1987,aklt_1988}, the minimum of the single-magnon dispersion shifts away from $k=\pi$, resulting in a change in curvature of the dispersion near this antiferromagnetic wavevector (see Fig.~\ref{fig:dsf_Haldane}). This is irreconcilable with the NL$\sigma$M prediction and corresponds to a negative $v^2$ in Eq.~\eqref{eq:nlsm_dispersion}. In the right panel of Fig.~\ref{fig:nlsm}, we compare the dynamic structure factor $S^{zz}(k, \omega)$ with the NL$\sigma$M result for the three-magnon continuum at momentum $k=\pi$ \cite{horton_1999,essler_2000} for several values of $\theta$. While they are qualitatively similar, the NL$\sigma$M curves have significantly stronger high-energy tails \cite{white_2008}, and the discrepancies become more pronounced when increasing $\theta$. Both shape and total spectral weight do not agree. Overall, the NL$\sigma$M predictions for the relevant quantities in the Haldane phase are unsatisfactory. 
\begin{figure*}[t]
\begin{center}
 \includegraphics[width=\textwidth]{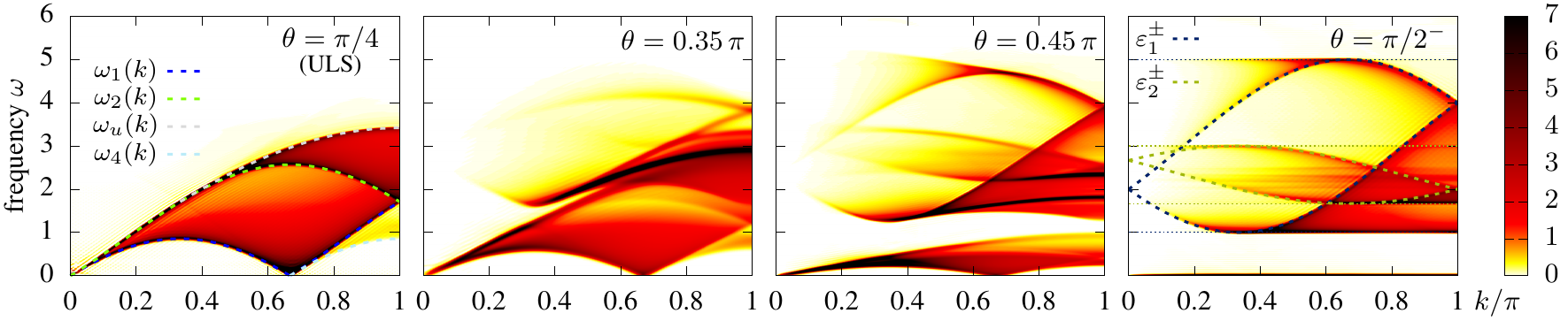}
 \end{center}
 \caption{Dynamic spin structure factors $S^{zz}(k, \omega)$ in the critical phase ($\pi/4\leq\theta<\pi/2$). Left: At the ULS point, we indicate exact continua boundaries from the nested Bethe ansatz solution. Center panels: With increasing $\theta$, the multi-soliton continua contract and higher-energy excitations emerge. Right: Just before the transition to the ferromagnetic phase, the soliton continua have collapsed onto the line $\omega=0$ and the higher energy features are captured by simple cosine dispersions \eqref{eq:elementaryBQ}.}
 \label{fig:dsf_critical}
\end{figure*}

While the NL$\sigma$M description fails quantitatively, it correctly predicts the presence of elementary magnon excitations with dispersion minimum at $k=\pi$ for an extended region in the Haldane phase. The stable single-magnon line and corresponding multi-magnon continua are clearly observed in the dynamic structure factors of Fig.~\ref{fig:dsf_Haldane}.
The exact shape of the excitations strongly depends on $\theta$. A lot of the features can be explained using a non-interacting approximation, where multi-magnon states are obtained by adding lattice momenta and energies $\varepsilon(k_i)$ of single-magnon states. This gives rise to boundaries and thresholds at jumps in the multi-magnon density of states as indicated in Fig.~\ref{fig:dsf_Haldane}. Jumps occur when group velocities $\mr{d}\varepsilon(k_i)/\mr{d} k_i$ agree for all magnons. Several of the threshold lines do not extend over the entire Brillouin zone, because the single-magnon states are only well defined down to a momentum $k_c(\theta)$, where $\varepsilon(k)$ enters a multi-magnon continuum, e.g., $k_c(0)\approx 0.23\pi$.
For small $\theta$, almost all features in $S$ correspond to such thresholds. See, for example, the lower boundaries of the two- and three-magnon continua and, for $\theta=0$, the structures at $(k,\omega)\approx (0.6\pi,3)$ and $(k,\omega)\approx (0.1\pi,5)$, which result from an interplay of jumps in the density of two- and three-magnon states. With increasing $\theta\gtrsim 0.1\pi$, the magnons interact more strongly and the non-interacting approximation cannot explain all structures anymore. At the AKLT point, for example, a sharp feature in the quadrupolar structure factor $S^{QQ}$ corresponds to an exactly known excited state with $S_{\mr{tot}}=2$, $k=\pi$, and energy $\omega=12/\sqrt{10}\approx 3.795$ \cite{moudgalya_2018}.

Tsvelik suggested a free Majorana field theory for the vicinity of the integrable Takhtajan-Babujan point $\theta=-\pi/4$ \cite{takhtajan_1982,babujian_1982,babujian_1983}. Surprisingly, we find that structure factors of that theory \cite{tsvelik_1990,essler_2000} deviate even stronger than the NL$\sigma$M results also near $\theta=-\pi/4$. This should be due to a neglect of current-current interactions.
Very recently, another alternative field-theoretic approach to the Haldane phase has been suggested \cite{kim_2020}. Instead of spin-coherent states, it uses an overcomplete basis of ``fluctuating'' MPS (fMPS) with bond dimension $D=2$, containing the AKLT ground state \cite{aklt_1987,aklt_1988}. Hence, the resulting Gaussian field theory works best around the AKLT point and reproduces the corresponding single-mode approximation for $\varepsilon(k)$ \cite{arovas_1988}. Figure~\ref{fig:nlsm} shows gaps and spin-wave velocities for the fMPS approach. It matches quite well around the AKLT point, but predicts the gap to close too early, at $\theta\approx 0.18\pi$
instead of at the Uimin-Lai-Sutherland (ULS) point $\theta=\pi/4$, and at $\theta\approx 0.04\pi$
instead of at the transition point $\theta=-\pi/4$ to the dimerized phase.

\section{Uimin-Lai-Sutherland point}
The transition from the Haldane phase to the critical phase occurs at the SU(3)-symmetric ULS point $\theta=\pi/4$. Here, the model can be solved using the nested Bethe ansatz \cite{uimin_1970,lai_1974,sutherland_1975}. The low-energy excitations are two types of soliton-like particles with $\varepsilon_1(k_1)=\left(\frac{2}{3}\right)^{3/2}\pi\,[\cos(\frac{\pi}{3}-k_1)-\cos\frac{\pi}{3}]$ for $k_1\in[0,\frac{2\pi}{3}]$ and $\varepsilon_2(k_2)=\left(\frac{2}{3}\right)^{3/2}\pi\,[\cos\frac{\pi}{3}-\cos(\frac{\pi}{3}+k_2)]$ for $k_2\in[0,\frac{4\pi}{3}]$, respectively. They are always created in pairs \cite{sutherland_1975,johannesson_1986}. Note that a computation of dynamical correlation functions based on the nested Bethe ansatz has not yet been achieved for this model. While recent work \cite{belliard_2012,wheeler_2013} has addressed the computation of scalar products of Bethe vectors, a single determinant representation has not yet been found. Hence, in the left panel of Fig.~\ref{fig:dsf_critical}, we show the numerical result for the dynamic structure factor $S^{zz}(k, \omega)$ and the boundaries of the relevant multi-soliton continua, which agree precisely with the main features. The line $\omega_1(k)$ indicates the lowest energy of a two-soliton excitation with total momentum $k$, the two-soliton density of states doubles at $\omega_2(k)$, and $\omega_u(k)$ is the upper boundary of the two-soliton continuum. In addition, a multi-particle continuum with less spectral weight can be found in the momentum range $k\in[\frac{2\pi}{3},\pi]$. Its lower bound $\omega_4(k)=\varepsilon_1(k-\frac{2\pi}{3})$ corresponds to four-soliton states.

\section{The critical phase}
As we increase $\theta$ starting from $\pi/4$, the soliton continua remain visible in the dynamic structure factor, but contract to lower energies as shown in Fig.~\ref{fig:dsf_critical}. In addition, further excitations emerge at higher energies.
The contraction of the continua can be explained by a field-theoretical description that is valid in the vicinity of the ULS point. In this region, the Hamiltonian can be mapped to a level-one SU(3) Wess-Zumino-Witten (WZW) model (action $\mc{A}_{\mr{SU}(3)_1}$), a conformal field theory with central charge $c=2$ and certain marginal perturbations \cite{itoi_1997}. As a function of $\theta$, the overall action can be written as
\begin{equation} \label{eq:action_marginal_terms}
	\mc{A}_\theta = \cos\theta \big[\mc{A}_{\mr{SU}(3)_1} + g_1(\theta) \mc{A}_1 + g_2(\theta) \mc{A}_2 \big].
\end{equation}
The first marginal term $\mc{A}_1$ describes an SU(3)-symmetric current interaction, which arises from constraining the dimension of the local Hilbert space and from a Gaussian integration over fluctuations of a mean-field variable \cite{itoi_1997}. The second marginal term $\mc{A}_2$ corresponds to the SU(3)-symmetry breaking Hamiltonian term $\hH_\theta-\hH_{\pi/4}$ with coupling $g_2\propto \tan\theta - 1$, where $g_2=0$ corresponds to the SU(3)-symmetric ULS point.
\begin{figure}[t]
 \begin{center}
  \includegraphics[width=0.95\columnwidth]{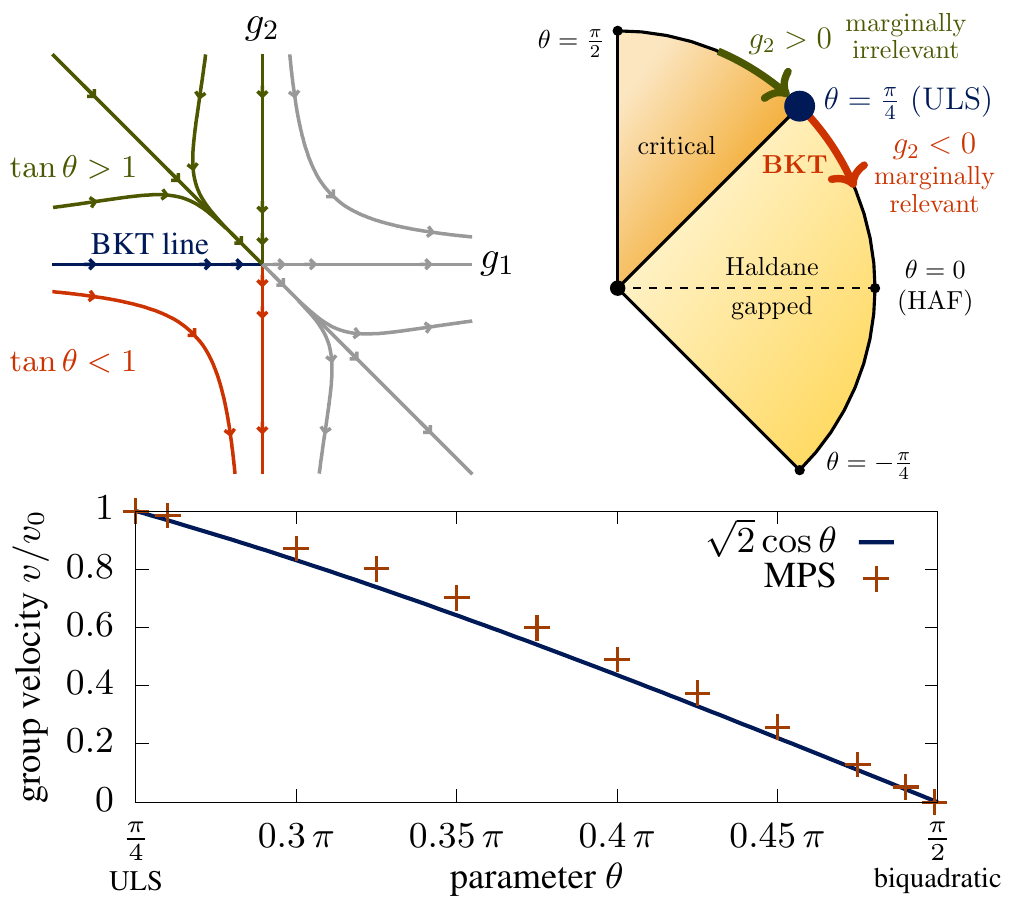}
 \end{center}
 \caption{Top: RG flow of the marginal terms in the field-theoretical description \eqref{eq:action_marginal_terms} in the vicinity of the ULS point \cite{itoi_1997} (left) and its relation to the phase diagram of $\hH_\theta$ (right). Bottom: Comparison of the $k=0$ group velocity extracted from the MPS simulations to the field-theoretical prediction. $v_0 = \sqrt{2}\pi/3$ is the exact group velocity at the ULS point.}
 \label{fig:contraction}
\end{figure}

Figure~\ref{fig:contraction} displays trajectories of the renormalization group (RG) flow for the couplings $g_1$ and $g_2$ of the marginal perturbations \cite{itoi_1997}. A comparison with the exact Bethe ansatz solution at the ULS point shows that the physically relevant trajectories start with $g_1 \leq 0$. In this regime, the term $\mc{A}_1$ is always marginally irrelevant and leads only to logarithmic finite-size corrections. Depending on the initial value of $g_2$, we have to distinguish two types of trajectories. For $g_2 < 0$ ($\theta < \pi/4$), the term $\mc{A}_2$ becomes marginally relevant, leading to a Berezinskii-Kosterlitz-Thouless (BKT) transition. Here, the model is asymptotically free with a slow exponential opening of the Haldane gap. For $g_2 \geq 0$ ($\theta \geq \pi/4$), the term $\mc{A}_2$ is marginally irrelevant and the RG flow approaches the only fixed point $g_1^* = g_2^* = 0$. Hence, the low-energy physics of this regime is described by the same field theory as the ULS point, corresponding to the presence of the extended critical phase.
Furthermore, the prefactor $\cos\theta$ in the action \eqref{eq:action_marginal_terms} explains the contraction of the multi-soliton continua for increasing $\theta$ as observed in Fig.~\ref{fig:dsf_critical}. Figure~\ref{fig:contraction} compares the numerically obtained group velocities in the critical phase to the field-theoretic prediction, finding very good agreement.

\section{Elementary excitations for \texorpdfstring{$\theta\to\pi/2^-$}{theta to pi/2}}
With increasing $\theta$, further higher-energy features emerge. To understand them, let us focus on the limit $\theta\to\pi/2^-$ (right panel in Fig.~\ref{fig:dsf_critical}). The low-energy continua have collapsed onto the line $\omega=0$ and we observe that the new dispersive excitations at higher energies can be described by intriguingly simple dispersion relations
\begin{subequations}\label{eq:elementaryBQ}
\begin{align}
	\varepsilon^{\pm}_1(k) &=3 + 2\cos(\pm k - 4\pi/3)\quad \text{and}\\
	\varepsilon^{\pm}_2(k) &= 7/3 + 2/3 \cos(\pm k -\pi/3).
\end{align}
\end{subequations}
Additional structures are constant-energy lines that appear at the minima and maxima of $\varepsilon^{\pm}_{1,2}(k)$, bounding corresponding excitation continua. The states in these continua can be explained as combinations of one of the massive excitations with one of the $\omega=0$ excitations with arbitrary momentum $k$.

To characterize the nature of the dispersive features, in particular, to assess whether they are due to elementary one- or two-particle excitations, we compare subsystem density matrices for the perturbed time-evolved state $|\psi(t)\ket \propto e^{-i\hH t}\hB_0|\psi\ket$ and the ground state $|\psi\ket$. Let us define block $\mc{A}$ as the left part of the spin chain up to but excluding the central site $x=0$ on which the perturbation is applied, and let us call the remainder of the system $\mc{B}$. Reduced density matrices for block $\mc{A}$ are obtained by a partial trace over the degrees of freedom of block $\mc{B}$, and we define $\hsigma_\mc{A}(t) := \Tr_\mc{B} |\psi(t)\ket\bra\psi(t)|$ and $\hrho_\mc{A} := \Tr_B |\psi\ket\bra\psi|$.
To quantify how similar the perturbed time-evolved states and the ground state are on block $\mc{A}$, we employ the block fidelity
\begin{equation} \label{eq:subsystem-fidelity}
	F_\mc{A}(t) := \left[ \Tr \sqrt{\sqrt{\hrho_\mc{A}}\, \hsigma_\mc{A}(t) \sqrt{\hrho_\mc{A}}} \right]^2. 
\end{equation}
For elementary single-particle excitations, we expect half of the weight of $|\psi(t)\ket$ to describe a left-moving particle. In this component, the state of the left subsystem is orthogonal to the ground state; hence it does not contribute to $F_\mc{A}(t)$. The other half describes a particle traveling to the right. On subsystem $\mc{A}$, this component looks like the ground state. We therefore expect $F_\mc{A}(t)$ to approach $1/2$ for large times. For elementary two-particle excitations, the wavefunction contains components describing one particle traveling to the left and one traveling to the right. There can be additional components with both particles traveling in the same direction. Only components where both particles travel to the right will contribute to $F_\mc{A}(t)$, which should hence approach a value significantly below $1/2$.
\begin{figure}[t]
 \begin{center}
  \includegraphics[width=0.96\columnwidth]{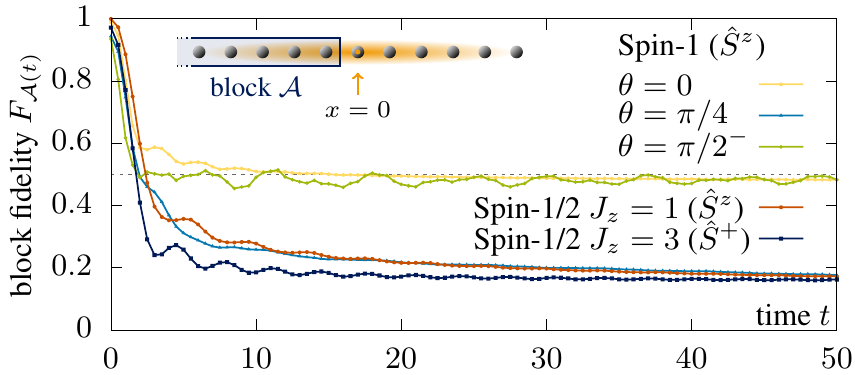}\vspace{-1.5em}
 \end{center}
 \caption{Evolution of block fidelities \eqref{eq:subsystem-fidelity} for different systems and perturbation operators $\hB$ as indicated in brackets. We show the spin-1 Heisenberg chain [$\theta=0$ in Eq.~\eqref{eq:H_blbq}] as an example for elementary one-particle excitations and want to characterize excitations at the biquadratic point $\theta=\pi/2^-$. Examples for elementary two-particle excitations include isotropic and anisotropic spin-$1/2$ XXZ chains and the spin-1 chain \eqref{eq:H_blbq} at the ULS point $\theta=\pi/4$.}
 \label{fig:block_fidelities}
\end{figure}

Figure~\ref{fig:block_fidelities} shows fidelities $F_\mc{A}(t)$ for several models. We include isotropic and anisotropic spin-$1/2$ XXZ chains, and the bilinear-biquadratic spin-1 chain \eqref{eq:H_blbq} at the ULS point $\theta=\pi/4$. For these three examples, we know that the dynamics is dominated by elementary two-particle excitations \cite{bethe_1931,cloizeaux_1962,cloizeaux_1966,yamada_1969,bougourzi_1996,bougourzi_1998,sutherland_1975,johannesson_1986}. As expected, $F_\mc{A}(t)$ converges to a small value significantly below $1/2$. For the spin-1 antiferromagnetic chain, where the dynamics is dominated by the single-magnon excitations, we confirm that the fidelity converges to $\approx 1/2$. The small deviation can be attributed to the contribution of multi-magnon excitations with relatively small spectral weight. For the spin-1 chain \eqref{eq:H_blbq} at $\theta=\pi/2^-$, we find that the block fidelity approaches $\approx 1/2$. This is a strong indication that the observed dispersive features in the dynamic structure factor in the right panel of Fig.~\ref{fig:dsf_critical} are due to elementary one-particle excitations.

\section{Temperley-Lieb chain and integrability}\vspace{-0.25em}
The simple functional form of the dispersions \eqref{eq:elementaryBQ} suggests that an exact solution is possible for $\theta=\pi/2^-$. At the purely biquadratic point $\theta=\pi/2$, the Hamiltonian is in fact frustration free and can be expressed as a sum of bond-singlet projectors $\hP_{i,i+1}$ such that $\hH_{\pi/2} = \sum_i (1 + 3\hP_{i,i+1})$. The groundstate space is exponentially large, containing all states without bond singlets. The projectors $\{\hP_{i,i+1}\}$ obey a Temperley-Lieb algebra \cite{temperley_1971,barber_1989}, which implies integrability of the model, and a corresponding generalization of the coordinate Bethe ansatz has been found \cite{koeberle_1994}. Starting from a ferromagnetic reference state, the $\hH_{\pi/2}$ eigenstates can be constructed by creating two types of pseudo-particles and adding so-called impurities. 
For $\theta=\pi/2^-$, an infinitesimal bilinear term $\sim\sum_i \hSvec_i \cdot \hSvec_{i+1}$ resolves the groundstate degeneracy. In terms of the Bethe ansatz, the resulting $\theta=\pi/2^-$ ground state is a specific linear combination of $\theta=\pi/2$ ground states containing a complex array of impurities and pseudo-particles. Unfortunately, the Bethe ansatz solution in its current form does not give access to this ground state. Hence, analytically deriving the dispersion relations \eqref{eq:elementaryBQ} remains an open problem. These massive excitations need to involve one bond singlet and, thus, $\varepsilon_{1,2}^\pm(k)\geq 1$.

\section{Conclusion}
We have explored the low-energy physics of isotropic spin-1 chains. The MPS algorithm \cite{binder_2018} allowed us to compute precise dynamic structure factors, even in the highly entangled critical phase with $c=2$. We have found that the NL$\sigma$M and Majorana field theories fail to capture the influence of the biquadratic term and provide only a rather unsatisfactory description for the Haldane phase. While an interpretation in terms of non-interacting magnons explains most features for small $\theta$, magnon interactions are quite important around and beyond the AKLT point, and a better field-theoretical understanding would be very valuable. In the critical phase, we observed and explained the contraction of the two-particle continua from the ULS point, finding agreement with field theory arguments. In addition, we have discovered new excitations at higher energies, which we have characterized to be of elementary one-particle type. For $\theta\to\pi/2^-$, their dispersion relations approach intriguingly simple forms. We hope that this observation will stimulate further research, possibly extending Bethe ansatz treatments for the integrable Temperley-Lieb chain.

\begin{acknowledgments}
We gratefully acknowledge helpful discussions with Fabian H.\ L.\ Essler, Israel Klich, Arthur P.\ Ramirez, and Matthias Vojta,
and support through US Department of Energy grant DE-SC0019449.
\end{acknowledgments}

\appendix

\section{Mapping to the non-linear sigma model} \label{sec:mapping_to_NLSM}
%
In this appendix, we explicitly show the calculations for the mapping of the bilinear-biquadratic spin-1 model \eqref{eq:H_blbq}
to the non-linear sigma model (NL$\sigma$M), complementing the discussion in the main text. We use a path-integral description based on spin-coherent states as, e.g., described in Ref.~\cite{fradkin_2013}, and show how the derivations need to be modified due to the presence of the biquadratic term.

\subsection{Path integral with spin-coherent states} \label{ssec:path_integral_spins}
%
For a single spin-$S$ with $\hS^z$ eigenbasis $\{|S;M\ket\}$, we define coherent states $|\vn\ket$ parametrized by unit vectors $\vn$. They obey $(\hSvec\cdot\vn)|\vn\ket=|\vn\ket$ and can be obtained by rotating the state with maximum $\hS^z$ quantum number by an angle $\chi$,
\begin{equation} \label{eq:spin_coherent_states}
 |\vn\ket := e^{i\chi \frac{(\vez \times \vn)}{\vert\vez \times \vn\vert}\cdot\hSvec} |S;S\ket,
\end{equation}
where $\vn\cdot\vez=\cos\chi$ and $\vez$ is the unit vector along the $z$ axis. These states can be used to derive a path integral representation for spin systems \cite{fradkin_2013}. 

Consider a spin chain 
\begin{equation*}
 \hH = \sum_i \hh_i
\end{equation*}
with nearest-neighbor interactions $\hh_i$ acting on sites $i$ and $i+1$ where, in the case of the bilinear-biquadratic chain,
\begin{equation}\label{eq:H_blbq2}
 \hh_i(\theta) \equiv  \cos\theta(\hSvec_i\cdot\hSvec_{i+1}) + \sin\theta(\hSvec_i\cdot\hSvec_{i+1})^2.
\end{equation}
Starting from the partition function in imaginary time, $Z = \Tr e^{-\beta\hH}$, one follows the usual procedure of discretizing time, $\beta = N\tau$, and inserting resolutions of the identity in terms of the states \eqref{eq:spin_coherent_states} for each intermediate time point and each lattice site $i$. This leads to the formal expression
\begin{equation}
 Z = \lim_{\substack{N\to\infty \\ N\tau = \beta}} \int\!\mc{D}[\{\vn_i\}]\, e^{-\mc{S}[\{\vn_i\}]}.
\end{equation}
Here, $\mc{D}[\{\vn_i\}]$ is an appropriate measure for the integral over the collection of smooth individual unit-vector paths $\vn_i(t)$ with periodic boundary conditions $\vn_i(0) = \vn_i(\beta)$. For the case of nearest-neighbor interactions, one can show that the Euclidean action $\mc{S}$ takes the form
\begin{align*}
 \mc{S}[\{\vn_i\}] = &-iS\sum_i\mc{S}_{\mr{WZ}}[\vn_i(t)]\\
  &+ \sum_i \int_0^\beta\!\mr{d}t\,\bra\vn_i(t),\vn_{i+1}(t)|\hh_i|\vn_i(t),\vn_{i+1}(t)\ket,
\end{align*}
where $|\vn_i,\vn_{i+1}\ket \equiv |\vn_i\ket \otimes |\vn_{i+1}\ket$ denotes the tensor product of two spins on neighboring sites. The first term is the sum of Wess-Zumino terms for individual spins, where $\mc{S}_{\mr{WZ}}[\vn_i(t)]$ is given by the total area of the cap on the unit sphere bounded by the (closed) trajectory $\vn_i(t)$.

\subsection{Evaluating the matrix element}
%
In order to obtain the action for the spin-1 model \eqref{eq:H_blbq} as a function of $\theta$, we need to evaluate the matrix element
\begin{align*}
 \bra\vn_i,\vn_{i+1}|&\hh_i(\theta)|\vn_i,\vn_{i+1}\ket\\
 &=\ \ \cos\theta \bra\vn_i,\vn_{i+1}|(\hSvec_i\cdot\hSvec_{i+1})|\vn_i,\vn_{i+1}\ket \\
 &\quad+ \sin\theta \bra\vn_i,\vn_{i+1}|(\hSvec_i\cdot\hSvec_{i+1})^2|\vn_i,\vn_{i+1}\ket.
\end{align*}
Evaluating the bilinear term is straightforward and yields
\begin{equation} \label{eq:matrix_element_bilinear}
 \bra\vn_i,\vn_{i+1}|(\hSvec_i\cdot\hSvec_{i+1})|\vn_i,\vn_{i+1}\ket = \vn_i\cdot\vn_{i+1}.
\end{equation}
For the evaluation of the biquadratic term, let
\begin{equation*}
 |\vn(\chi)\ket := e^{i\chi\hS^y}|S=1;M=1\ket.
\end{equation*}
Then $|\vn(0)\ket = |1;1\ket$, and we consider the matrix element
\begin{equation}\label{eq:fchi}
 f(\chi) := \bra\vn(0),\vn(\chi)|(\hSvec_1\cdot\hSvec_2)^2|\vn(0),\vn(\chi)\ket.
\end{equation}
Writing the operator in the form $(\hSvec_1\cdot\hSvec_2)^2 = \big(\frac{1}{2}\hS^+_1\hS^-_2 + \frac{1}{2}\hS^-_1\hS^+_2 + \hS^z_1\hS^z_2\big)^2$, one obtains nine terms from expanding the square, and it is straightforward to see that only the two terms $(\hS^z_1)^2(\hS^z_2)^2$ and $\frac{1}{4}(\hS^+_1\hS^-_1)(\hS^-_2\hS^+_2)$ yield non-zero contributions in Eq.~\eqref{eq:fchi}. Therefore,
\begin{equation*}
\begin{split}
 f(\chi) &= \bra\vn(0)|(\hS^z)^2|\vn(0)\ket \bra\vn(\chi)|(\hS^z)^2|\vn(\chi)\ket\\
 &\quad+ \frac{1}{4} \bra\vn(0)|\hS^+ \hS^-|\vn(0)\ket \bra\vn(\chi)|\hS^- \hS^+|\vn(\chi)\ket \\
 &= \bra\vn(\chi)|(\hS^z)^2|\vn(\chi)\ket + \frac{1}{2} \bra\vn(\chi)|\hS^- \hS^+|\vn(\chi)\ket.
 \end{split}
\end{equation*}
Note that $\hS^-\hS^+ = \hSvec^2 - (\hS^z)^2 - \hS^z$, and we can easily read off $\bra\vn(\chi)|\hSvec^2|\vn(\chi)\ket=2$ as well as $\bra\vn(\chi)|\hS^z|\vn(\chi)\ket = \cos\chi$, because we have $S=1$ and $\hSvec$ transforms like a vector under rotations. To evaluate the remaining matrix element $\bra\vn(\chi)|(\hS^z)^2|\vn(\chi)\ket$, we expand the rotated state $|\vn(\chi)\ket$ in the $\hS^z$ eigenbasis $\{|1;M\ket\}$,
\begin{align*}
 |\vn(\chi)\ket &= \sum_{M'=-1}^1 |1;M'\ket\bra 1;M'| e^{i\chi\hS^y}|1;1\ket\\
  &= \frac{1 + \cos\chi}{2} |1;1\ket + \frac{\sin\chi}{\sqrt{2}} |1;0\ket + \frac{1-\cos\chi}{2} |1;-1\ket,
\end{align*}
where the coefficients are entries of the representation matrix for spin-$1$ rotations [Wigner (small) $d$ matrix]. Hence,
$\bra\vn(\chi)|(\hS^z)^2|\vn(\chi)\ket = \frac{1}{4}(1 + \cos\chi)^2 + \frac{1}{4} (1-\cos\chi)^2
 = \frac{1}{2} + \frac{1}{2}\cos^2\chi$.
Putting everything together, we obtain $f(\chi) = \frac{5}{4} - \frac{1}{2} \cos\chi + \frac{1}{4}\cos^2\chi$.
As $(\hSvec_1\cdot\hSvec_2)^2$ transforms as a scalar under rotations, the matrix element depends only on the angle between the two spin-coherent states. Thus, the calculation generalizes to any two states $|\vn_1, \vn_2\ket$, and we can replace $\cos\chi$ by $\vn_1\cdot\vn_2$, obtaining
\begin{equation*}
 \bra\vn_1,\vn_2|(\hSvec_1\cdot\hSvec_2)^2|\vn_1,\vn_2\ket = \frac{5}{4} - \frac{1}{2} \vn_1\cdot\vn_2 + \frac{1}{4} (\vn_1\cdot\vn_2)^2.
\end{equation*}
Combining this result with Eq.~\eqref{eq:matrix_element_bilinear}, we arrive at the matrix element of the Hamiltonian interaction \eqref{eq:H_blbq2}:
\begin{align} 
 \bra\vn_i,\vn_{i+1}|&\hh_i(\theta)|\vn_i,\vn_{i+1}\ket\nonumber\\
 =& \frac{5}{4} \sin\theta + \Big(\cos\theta - \frac{1}{2}\sin\theta \Big) (\vn_i\cdot\vn_{i+1})\nonumber \\\label{eq:matrix_element_nlsm}
 &+ \frac{1}{4} \sin\theta (\vn_i\cdot\vn_{i+1})^2.
\end{align}
Note that
  $(\vn_i+\vn_{i+1})^2 = \vn_i^2 + 2\vn_i\cdot\vn_{i+1} + \vn_{i+1}^2 = 2\vn_i\cdot\vn_{i+1} + 2$, and
  $(\vn_i+\vn_{i+1})^4 = (2\vn_i\cdot\vn_{i+1} + 2)^2 = 4(\vn_i\cdot\vn_{i+1})^2 + 8 \vn_i\cdot\vn_{i+1} + 4$,
such that
\begin{equation*}
 \begin{split}
  \vn_i\cdot\vn_{i+1} &= \frac{1}{2}(\vn_i+\vn_{i+1})^2 + \mr{const}, \\
  (\vn_i\cdot\vn_{i+1})^2 &= \frac{1}{4} (\vn_i+\vn_{i+1})^4 - (\vn_i+\vn_{i+1})^2 + \mr{const}.
 \end{split}
\end{equation*}
Inserting this into Eq.~\eqref{eq:matrix_element_nlsm} yields for the matrix element, up to an irrelevant additive constant:
\begin{multline}
 \bra\vn_i,\vn_{i+1}|\hh_i(\theta)|\vn_i,\vn_{i+1}\ket = \frac{\cos\theta - \sin\theta}{2} (\vn_i+\vn_{i+1})^2\\
 + \frac{\sin\theta}{16} (\vn_i+\vn_{i+1})^4 + \mr{const}.
\end{multline}
Then, as a function of $\theta$, the action for the bilinear-biquadratic spin-1 chain \eqref{eq:H_blbq2} is given by
\begin{align}
 \mc{S}_\theta[\{\vn_i\}] = &-i \sum_i\mc{S}_{\mr{WZ}}[\vn_i(t)] \nonumber \\
 &+ \!\int_0^\beta\!\!\!\mr{d}t\,\sum_i \Big[\frac{\cos\theta - \sin\theta}{2}(\vn_i(t)+\vn_{i+1}(t))^2 \nonumber \\  \label{eq:action_blbq}
 &\qquad\qquad\ \ + \frac{\sin\theta}{16}  (\vn_i(t)+\vn_{i+1}(t))^4 \Big].
\end{align}
The special case $\theta=0$ corresponds to the spin-1 Heisenberg antiferromagnet, for which the original derivation was done \cite{haldane_1983_pla,haldane_1983_prl,affleck_1989}.

\subsection{Continuum limit and non-linear sigma model mapping}
%
In the next steps of the derivation, we follow the same approach that was taken for the Heisenberg antiferromagnet \cite{haldane_1983_pla,haldane_1983_prl,affleck_1989,fradkin_2013}. It is reasonable to expect staggered short-range order for the spin field $\vn$, and the most relevant low-energy modes should be ferromagnetic and antiferromagnetic fluctuations. Hence, we can choose an ansatz that separates these relevant degrees of freedom,
\begin{equation} \label{eq:field_separation_nlsm}
 \vn_i = (-1)^i \sqrt{1-a^2\vl_i^2}\, \vm_i + a\vl_i,
\end{equation}
where $a$ is the lattice spacing, and we have the constraints $\vm_i^2 = 1$ and $\vm_i \cdot \vl_i = 0$. Here, $\vm_i$ and $\vl_i$ are slowly varying, which allows us to take the continuum limit $a\to 0$. We can write $\vm_{i+1} \approx \vm_i + a (\partial_x \vm_i)$ and similarly for $\vl_i$. When inserting the ansatz \eqref{eq:field_separation_nlsm} into the action \eqref{eq:action_blbq}, we only need to keep terms to the lowest order in $a$. For the first term, this yields
\begin{multline*}
 \frac{1}{2}(\vn_{i-1}+\vn_i)^2 + \frac{1}{2}(\vn_i+\vn_{i+1})^2 \\ = a^2 \left((\partial_x \vm_i)^2 + 4 \vl_i^2 \right) + \mc{O}(a^3),
\end{multline*}
where we have grouped two neighboring interaction terms together to take advantage of the cancellation of additional terms. Correspondingly, the contributions from the second term $(\vn_i+\vn_{i+1})^4$ will be of the order $\mc{O}(a^4)$. Hence, the second term can be ignored in the continuum (low-energy) limit, and the effective action has the same form as in the case of the Heisenberg antiferromagnet ($\theta=0$). The only change due to the biquadratic term is an effective rescaling of the coupling in the form $J(\theta) = \cos\theta - \sin\theta$. Thus, the remaining steps in the derivation for the mapping to the NL$\sigma$M are identical to the case of the Heisenberg antiferromagnet.

After taking the continuum limit for the Wess-Zumino terms as well, one can integrate out the fluctuations in the field $\vl$, which yields an effective action
\begin{equation} \label{eq:action_afm_chain}
 \mc{S}[\vm] = \!\iint\!\mr{d}x\,\mr{d}t\,\frac{1}{2g}\left(v(\theta)\,(\partial_x \vm)^2 + \frac{(\partial_t \vm)^2}{v(\theta)} \right) + i\phi \mc{Q}[\vm],
\end{equation}
where we have introduced the coupling constant $g=2/S$, the spin wave velocity $v(\theta)=2aJ(\theta)S$, and the topological angle $\phi=2\pi S$. The second term contains the topological charge or winding number of the field configuration
\begin{equation*}
 \mc{Q}[\vm] = \frac{1}{8\pi} \iint\!\mr{d}x\,\mr{d}t\,\epsilon_{ij} \vm\cdot(\partial_i \vm \times \partial_j \vm) \in \mathbb{Z}.
\end{equation*}
Note that for integer spin $S$, the imaginary part $\phi \mc{Q}[\vm]$ in Eq.~\eqref{eq:action_afm_chain} is always an integer multiple of $2\pi$, such that it does not affect the physics. In this case, the model is described by the first term, which is the standard $O(3)$ non-linear sigma model (NL$\sigma$M). For half-integer spin, however, the contributions to the path integral of configurations with an odd winding number $\mc{Q}$ are weighted by a factor $-1$. This leads to fundamentally different physics, which is at the core of Haldane's conjecture \cite{haldane_1983_pla,haldane_1983_prl,affleck_1989}. While antiferromagnetic chains with integer spin are gapped, those with half-odd-integer spin are gapless.

In conclusion, the low-energy physics of the bilinear-biquadratic spin-1 chain should be described by the NL$\sigma$M, which predicts an excitation gap $\Delta(\theta) \propto J(\theta) e^{-\pi S}$ and a dispersion $\varepsilon(k) = \sqrt{\Delta^2 + v^2 (k-\pi)^2}$ for the single-magnon line near $k=\pi$. In the main text, we are testing the dependence of the gap and the spin-wave velocity on the Hamiltonian parameter $\theta$, for which we summarize the NL$\sigma$M predictions:
\begin{equation}
\begin{split}
 \Delta(\theta) &\propto (\cos\theta - \sin\theta)\quad \text{and}\\
    v^2(\theta) &\propto (\cos\theta - \sin\theta)^2.
\end{split}
\end{equation}

\newpage
\bibliographystyle{apsrev4-1_with_titles}

\end{document}